# Global structure of the Local Universe according to 2MRS survey

## Tekhanovich D.I.[1] and Baryshev Yu.V.[1*]

*[1]Astronomical Department, Saint Petersburg State University, Saint-Petersburg, Russia*

*\* yubaryshevl@mail.ru*

**Abstract** We report the results of a statistical analysis of the space distribution of galaxies within distances about 300 Mpc using the 2MRS catalog, which contains redshifts of 43533 galaxies of the 2MASS all-sky IR survey. Because of the unique features of the 2MRS survey, such as its 90% sky coverage, galaxy selection in the IR, the complete incorporation of the old stellar population of galaxies, weakness of the dust extinction effects, and the smallness of the k- and e-corrections allowed us to determine the statistical properties of the global distribution of galaxies in the Local Universe. We took into account the main methodological factors that distort the theoretically expected relations compared to those actually observed. We construct the radial galaxy number counts N(R), SL(R, r) statistics, and the complete correlation function (conditional density) Γ(r) for volume-limited (VL) galaxy samples. The observed conditional density Γ(r) in the redshift space is independent of the luminosity of galaxies and has the form of a power-law function $\Gamma(r) \sim r^{-\gamma}$ with exponent $\gamma \sim 1.0$ over a large range scale-length spanning from 0.1 to 100 Mpc. We compare the statistical properties of the space distribution of galaxies of the 2MRS catalog with the corresponding properties of simulated catalogs: stochastic fractal distributions and galaxies of the Millennium catalog.

Keywords: cosmology, large-scale structure of the Universe, cosmology, observations

## 1. Introduction

Local Universe is the space region of radius $R < 300$ Mpc ($z < 0.1$) centered on the terrestrial observer. It is the most thoroughly studied part of the Universe which presents the result of the evolution of galaxies and systems of galaxies over the Hubble time. In particular, the actually observed space distribution of galaxies in the Local Universe can be used to test the predictions of the formation and evolution models of the large-scale structure of the Universe [1, 2].

Current studies of the Local Universe opened up a great variety of large scale structures — groups, clusters, voids, and superclusters, which are studied both as individual objects and as elements of the global distribution of galaxies [3–11]. The greatest structural features have the form of filaments, walls, and voids located between them, and their linear sizes can amount to several hundred Megaparsecs [7, 11, 12].

One of the most fundamental statistical properties of the general space distribution of galaxies, which includes complex observed structures (filaments, voids, shells, and walls), is the fractal dimension of the global structure as a whole. Fractal approach to the analysis of the distribution o galaxies was first used by [1, 13–16]. Detailed review of the history and prospects of the fractal approach to the large-scale distribution of galaxies see [17].

The fractal dimension $D = 3 - \gamma$ inferred from the complete correlation function in the redshift space is a directly determined quantity, which characterizes the spatial, kinematic, and dynamic state of the Local Universe. It can be estimated from the power-law slope $\gamma$ of the full correlation function $\Gamma(r) \sim r^{-\gamma}$ without invoking any a priori assumptions about the evolution of non-baryonic dark matter and its association with baryonic matter (galaxies) or the form of the distribution of peculiar velocities of galaxies. The account for the actual peculiar velocities of galaxies would require further development of redshift-independent methods for determining galaxy distances and performing time consuming observational programs aimed at mass measurement of such distances [18].

Note that stochastic fractal structures naturally arise in physics as a result of the dynamical evolution of complex systems. Physical fractals are discrete stochastic systems characterized by power-law correlation functions. In particular, fractal structures arise in turbulent flows [19]. Deterministic chaos of nonlinear dynamic systems is characterized by fractal dimension, which measures the strangeness degree of strange attractors [20]. Phase transitions and thermodynamics of self-gravitating systems are also characterized by the formation of fractal structures [21, 22], however, many important aspects of these studies so far remain undiscovered.

In this paper we report the results of our statistical analysis of the distribution of galaxies of the 2MRS catalog [23], which contains redshifts of 43533 galaxies of the all-sky 2MASS IR survey. We pay special attention to taking into account selection effects, which are inevitably present in observational data and which distort the theoretically expected relations compared with those directly observed. In particular, of importance are such properties of the surveys as the opening angle of the observed continuous convex region, limiting magnitude of the survey, total number of galaxies and the average distance between galaxies, the size of the biggest sphere fully immersed in the sample volume, and the number of independent spheres completely fitting inside the sample volume. One must also take into account the initial assumptions concerning the applicability of the mathematical models used for data interpretation, such as the use of continuous gas-dynamic quantities or discrete fractal structures, as well as the ergodicity and stationarity of real galaxy samples.

In Section 2 we describe the observational data employed and construct the VL samples required for the application of the statistical methods of analysis used in this study. Section 3 presents the results for the $N(R)$ and SL($R, r$) statistics. In Section 4 we analyze the full correlation function (conditional density) $\Gamma(r)$ and the results of its application to VL samples. In Section 5 we analyze similar statistical properties of simulated galaxy catalogs – stochastic fractal distributions and samples drawn from the Millennium. In Section 6 we discuss the results of a comparison of the actually observed distribution of 2MRS galaxies with the corresponding distributions for simulated catalogs and present the main conclusions.

## 2. Observational data

In this paper we use the 2MASS XSC Redshift Survey (2MRS) catalog [23], which represents the spectroscopic part o the biggest IR survey of Local Universe galaxies (2MASS). The 2MRS catalog contains redshifts and apparent infrared magnitudes for all galaxies with $Ks$ < 11.75 and color excesses $E(B - V) < 1.0$ in the $|b| > 5°$ region (for $l > 330°$ or $l < 30°$ the catalogue region has $|b| > 8°$ near the Galactic center) (Fig. 1). The 2MRS survey contains a total of 43533 galaxies and covers 90% of the sky out to the distance of 300 Mpc. Galaxies are selected in the infrared and therefore all interstellar-extinction related effects are one order of magnitude weaker than in the optical. Furthermore, in the infrared the redshift (*k-correction*) and evolution (*e-correction*) corrections proved to be insignificant inside the sample volume for the Local Universe.

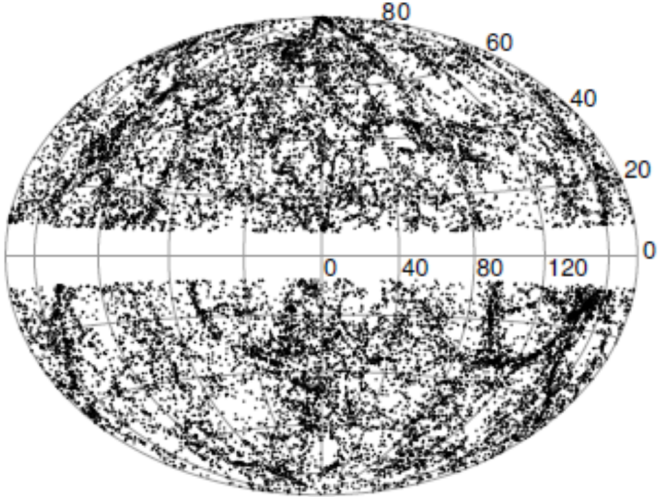

**Fig. 1.** *Distribution of 2MRS catalog galaxies in Galactic coordinates. Only galaxies in the 50 < R < 150 Mpc and |b| > 8 ∘ intervals are shown.*

The 2MRS survey is 97% complete down to the limiting magnitude of $Ks$ = 11.75. Hence the 2MRS catalog is a statistically uniform galaxy sample, which covers almost the entire sky and represents a "fair sample" of the space distribution of galaxies in the Local Universe. Note that the completeness of the 2MRS catalog naturally refers to the set of its fixed inner limits. In particular, the selection criteria for IR galaxies of the initial 2MASS survey imply limited visibility of low surface brightness galaxies and galaxies with predominantly blue population.

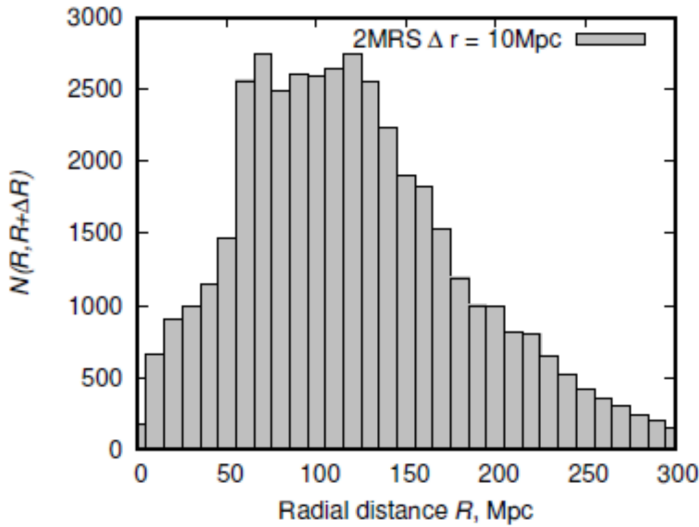

**Fig. 2.** *Radial distribution of 2MRS galaxies in ΔR = 10 Mpc thick layers.*

We used the 2MRS catalog data about the position (Galactic coordinates $l$, $b$), redshift ($z$, in the barycentric reference frame), and apparent $K$-band magnitude ($Ks$). When passing to Cartesian space coordinates we transformed the redshifts into metric distances $R$ using the following standard formula:

$$R(z) = \frac{c}{H} \int_{1/(1+z)}^{1} \frac{dy}{y\sqrt{\Omega_m/y + \Omega_L y^2}} \qquad (1)$$

where $H = 70$ km/s Mpc, $\Omega_m = 0.25$, and $\Omega_L = 0.75$. We then computed the absolute magnitudes of galaxies from their radial distances and apparent magnitudes by the formula:

$$M_k = K_s - 5\log\big(R(z)(1+z)\big) - 25 \qquad (2)$$

As we pointed out above, we compute the absolute magnitudes without applying the redshift and evolutionary corrections, because they are negligible on scales considered in our analysis. Small corrections for Galactic foreground extinction are already included in the 2MRS catalog.

**Table 1.** *Basic parameters of 2MRS VL samples: radial distance and Ks-band absolute magnitude limits ($R_{lim}$ and $M_{lim}$), the total number N of galaxies within these limits in the $b > 8°$ (S1) and $b < -8°$ (S2) hemispheres, the ratio of the number of galaxies to the sample volume, and the average distance to the nearest neighbor, $<R_{near}>$*

|         | $R_{\text{lim}}$, Mpc | $M_{\text{lim}}$ | $N$ | $N/V$, Mpc$^{-3}$ | $\langle R_{\text{near}} \rangle$, Mpc |
|---------|------|---------|------|-------------|---------|
| S1 VL0  | 50   | $-21.79$ | 1749 | $7.76e-03$  | 1.54    |
| S1 VL1  | 138  | $-24.02$ | 5567 | $1.17e-03$  | 3.11    |
| S1 VL2  | 200  | $-24.85$ | 3205 | $2.22e-04$  | 5.99    |
| S1 VL3  | 250  | $-25.37$ | 1566 | $5.56e-05$  | 10.02   |
| S2 VL0  | 50   | $-21.79$ | 900  | $3.99e-03$  | 1.95    |
| S2 VL1  | 138  | $-24.02$ | 5333 | $1.13e-03$  | 3.12    |
| S2 VL2  | 200  | $-24.85$ | 3466 | $2.40e-04$  | 5.75    |
| S2 VL3  | 250  | $-25.37$ | 1814 | $6.44e-05$  | 9.58    |

## 2.1. Volume Limited samples of the 2MRS catalog

The fixed limiting magnitude, used for selecting galaxies of the 2MRS survey, produces strong radial selection relative to the observer. To eliminate this selection effect, we used the standard approach involving the construction of volume-limited (VL) samples, which was employed by [15, 24, 25]. The observed absolute magnitude–radial distance relation (Fig. 3) is used to select the $M_{lim}$ and $R_{lim}$ values so that all objects brighter than some fixed luminosity would be included into the sample out to a certain distance.

We selected a total of four $M_{lim}$ and $R_{lim}$ combinations: $(-21.79, 50)$, $(-24.02, 138)$, $(-24.85, 200)$, and $(-25.37, 250)$. Based on these combinations, we constructed eight VL samples in the Northern and Southern hemisphere ($b > 8°$ and $b < -8°$, which are denoted as S1 and S2, respectively). We list the basic parameters of these samples in Table 1.

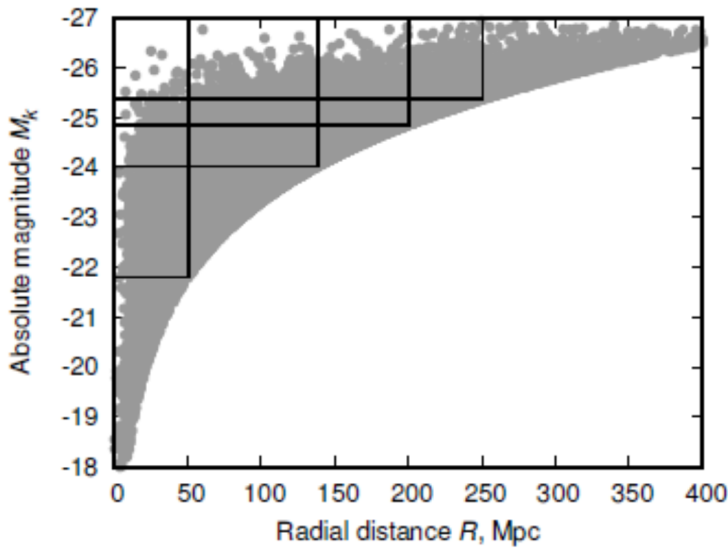

**Fig. 3.** *Distribution of 2MRS catalog galaxies in the absolute magnitude–heliocentric distance plane. The black lines show the boundaries of the VL samples used in this study.*

## 3. Radial counts and SL statistics of VL samples

According to the definition of discrete fractal distribution [13], the average number of objects in spheres circumscribed around points $p$ of this distribution should vary in accordance with a power law as a function of the radius $r$ of the spheres considered, i.e. is given by:

$< N(r) >_p = (r/r_0)^D$ . Exponent $D$ cannot exceed the dimension of the space where the distribution is located; it is referred to as fractal dimension and is the main characteristic of the fractal distribution. The definition of fractal dimension includes averaging over all points of the set, therefore counting points from a single center (the observer's position) may differ significantly from the theoretical fractal behavior $< N(r) >_p$. On the other hand, an analysis of radial distributions around the observer *(distance R)* allows one to coarsely estimate the behavior of the variation $N(R) \sim R^{\alpha}$ with minimum additional assumptions.

We analyzed radial distributions for different volume-limited (VL) samples and show the results of our counts for VL2 sample in Fig. 4. The radial distributions around the observer in different hemispheres differ at heliocentric distances out to 100 Mpc, whereas at greater distances these distributions appear more similar and close to $R^3$. It is remarkable that radial counts made around galaxies drawn from central regions of VL samples yield $N(R) \sim R^{\alpha}$ estimates that differ from that obtained from counts centered on the observer located in our Galaxy. Counts made on 40–80 Mpc scales for galaxies located at the center of the northern VL2 region yield $\alpha = 2.1$, whereas those made for the southern region yield $\alpha = 3.1$. The average exponent $\alpha$ is close to the estimate earlier found for KLUN galaxy sample [26]. Figure 4 shows large fluctuations in the behavior of radial galaxy counts made from the same observing point.

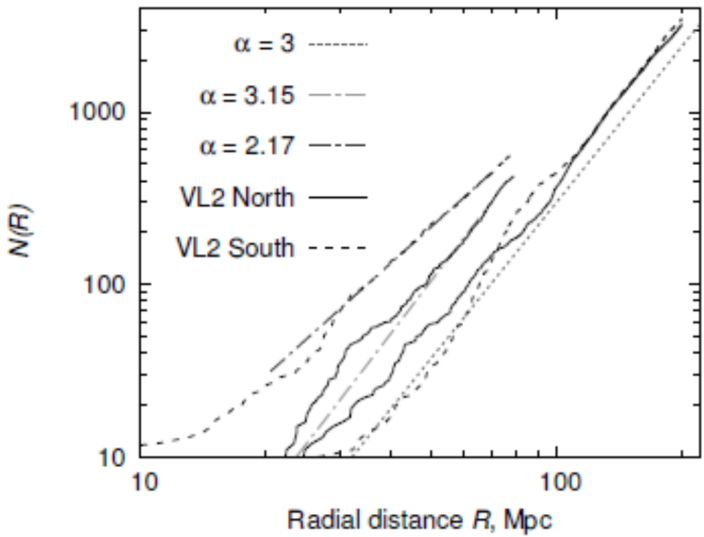

**Fig. 4.** *Example of galaxy counts in the VL2 sample. The solid and dashed lines show the counts made in the northern and southern hemisphere, respectively. For each hemisphere the counts were made both around our own Galaxy (out to 200 Mpc) and around the galaxy closest to the center of the sample.*

Sylos Labini et al. [27] proposed a new statistical property (scale–length (SL)) for non-uniform discrete distributions that characterizes fluctuations of the number of galaxies in test spheres as a function of the distance from the observer. Figure 5 shows the SL statistics: $N_i(R, r)$ is the number of points inside the sphere of radius $r$ centered on $i$-th point located at distance $R$ from the observer. SL statistics is presented for VL2 sample in the northern hemisphere and scale length $r = 30$ Mpc. Also shown is the variation of the average value $Ni(R, r = 30)$ as a function of $R_k$; we computed the average value in the $R_k - 15 < R_k < R_k + 15$ region for a set of $R_k$ values. The error bars correspond to the standard deviation in each bin.

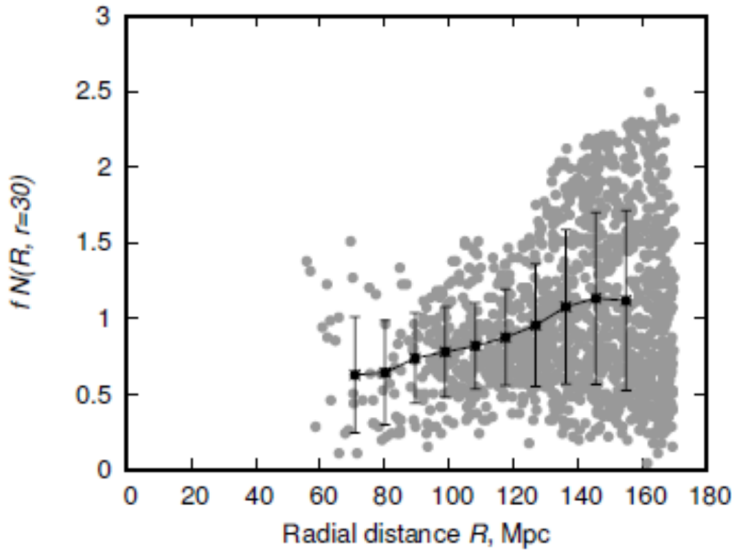

**Fig. 5.** *Normalized SL statistics for VL2 in the northern hemisphere. Also shown are the average $N_l(R, r = 30)$ values in $\Delta R = 30$ Mpc radial bins. Error bars correspond to standard deviation in each bin. The average increases with increasing distance.*

SL statistics also tests the fulfillment of the condition that is necessary for "self-averaging": angular and translational invariance of the sample and, in particular, the absence of any systematic trend of the number of points in spheres with distance $R$, which can be due to the presence of inhomogeneities with sizes exceeding the size of the sample [27]. Figure 5 shows that both the mean and the variance of the SL statistics increase with distance, implying the presence of large structures in the sample of 2MRS galaxies.

## 4. Complete correlation function for VL samples

### 4.1. General description

To study the correlation properties of the distribution of galaxies, we use the complete correlation function (conditional density). Pietronero [14] was the first to introduce this function to describe distributions of galaxies, and its detailed analysis was performed in the monograph by [16]. It was also used to analyze many catalogs of galaxies [15, 28, 29]. The complete correlation function of a point-wise stochastic set in $d$-dimensional space is given by the following formula:

$$\Gamma_d(\boldsymbol{r_{12}}) = \frac{< n(\boldsymbol{r_1})n(\boldsymbol{r_2}) >}{< n(\boldsymbol{r}) >} \qquad (3)$$

where $n(\mathbf{r}_i)$ is the particle number density inside volume $dV_i$ around point $i$ with the coordinates $\mathbf{r}_i$; and $\mathbf{r}_{12} = \mathbf{r}_1 - \mathbf{r}_2$, the vector of the distance between points 1 and 2, and $<x>$ is the ensemble average of $x$.

### 4.1.1. Integrated Conditional Density

The statistical estimate of the complete correlation function, which is also referred to as conditional density, for the galaxy sample considered is defined as

$$\Gamma_d(r) = \frac{1}{N_c(r)} \sum_{i=1}^{N_c(r)} \frac{N_i(r)}{V(r)} \qquad (4)$$

Where $N_i(r)$ is the number of points inside spherical volume $V_d(r)$ of the $d$-dimensional space circumscribed about $i$-th point and $N_c(r)$ is the number of centers of test spheres, i.e., the number of points around which this volume is circumscribed. It is important to bear in mind that averaging has to be performed without going beyond the initial distribution volume, and this restriction has important effect on the greatest scale lengths, i.e., on the greatest test spheres. This condition strongly restricts the scale lengths accessible for the analysis of galaxy distributions, because, strictly speaking, because to reliably compute the conditional density on some selected scale, we must analyze one order of magnitude greater continuous spherical region.

### 4.1.2. Statistical Error of Conditional Density

Formula (4) actually yields the sample average of the number of particles in all test spheres of the sample volume considered. If the sample variance of the mean is adopted as the formal measure of the error of conditional density then the error boxes would be smaller than the thickness of the corresponding lines in plots practically for all scale values. Of much greater importance is the allowance for various factors distorting the true form of the conditional density distribution of the space distribution of galaxies considered.

## 4.2. Factors That Distort the Estimate of the Full Correlation Function

When estimating the conditional density all points of the galaxy sample considered are assumed to be statistically equal. However, overt and hidden selection effects, which are inevitably present in observational data, distort the theoretically expected relations compared to those actually observed.

Of special importance is the effect of the limited volume of the galaxy sample considered. Because of this effect the behavior of conditional density on scales close to the sample size may change significantly when we analyze deeper samples. This is due to the fact that only spheres fitting completely inside the sample boundaries can be used for bona fide determination of conditional density, because the distribution of galaxies outside the sample volume is unknown. Note that in the Davis–Peebles method [30] the reduced correlation function $\xi(r)$ is computed assuming that the average value remains constant both inside and outside the sample volume. Because of this, staring from certain scale lengths the number of independent spheres used to determine the average value begins to decrease rapidly with scale considered or their centers cover increasingly smaller region and the number of independent (nonintersecting) test volume decreases even faster.

Another important distorting factor is the limited coverage of sky directions in the sample. Because of this effect there is no averaging over angle coordinates. That is why all-sky surveys yield significantly more reliable estimates for the parameters of the distributions of galaxies compared to surveys with limited sky coverage.

Furthermore, because of the discreteness of the sample and the finite number of points in the distribution the range of scale lengths that can be studied using the conditional density method has also a lower boundary. A natural estimate for the minimum scale length starting from which the estimate of conditional density becomes significant is given the distance to the nearest neighbor, $<R_{near}>$. The lower boundary of reliable estimation of conditional density is determined by the fraction of empty spheres in small-scale spheres.

In the general case, i.e., for arbitrary distribution, there are no analytical relations for determining the reliability of the conditional density estimate. An analysis of simulated fractal distributions suggests the following stringent criterion for the "reliability" of the determination of conditional density: the number of spheres $N_c(r)$ used for averaging should be equal to or greater than half the total number of points in the sample considered. We define this ratio of $N_c(r)$ to $N$ as $v(r) = N_c(r)/N$. It is evident that the geometry of the sample used contributes significantly to the variation of $v(r)$ and to the error of the conditional density estimate. Thus if we consider a uniformly filled sphere of radius $R = 300$ Mpc then the condition $v(r) > 0.5$ fulfilled by scale lengths shorter than $r_{max} \approx 61$ Mpc, whereas this maximum scale length reduces to $r_{max} \approx 38$ Mpc for half-sphere and to $r_{max} \approx 23$ Mpc for 1/8 sphere.

Thus the **principle sample parameters** to be controlled in the correlation analysis of the space distribution of galaxies include:

- *angular size of the observed continuous sky region inside which the galaxy sample is defined;*
- *limiting magnitude of the survey, the upper and lower cutoff of the luminosity function, effects of the large and the small volume-size at large and small distances from the observer;*
- *total number of galaxies and average distance between them; size of the maximum sphere that can be fully immersed in the sample volume, the number of spheres as a function of scale-length, and the number of independent spheres;*
- *initial assumptions of the statistical data analysis method employed and, in particular, the need of the fulfillment of the condition of representativity of statistical average values ("self-averaging") [27].*

### 4.3. Three-Dimensional Conditional Density

We estimated the conditional density for VL-samples of 2MRS galaxies (4). Figure 6 shows the result for eight VL samples (numbering from 1 to 4 corresponds to samples VL0 – VL3). The power law of conditional density $\Gamma(r) \sim r^{-\gamma}$ is the same for all samples and its exponent $\gamma$ is close to 1 in the scale-length range covered by the corresponding VL sample. This slope corresponds to the fractal dimension of $D \approx 2$.

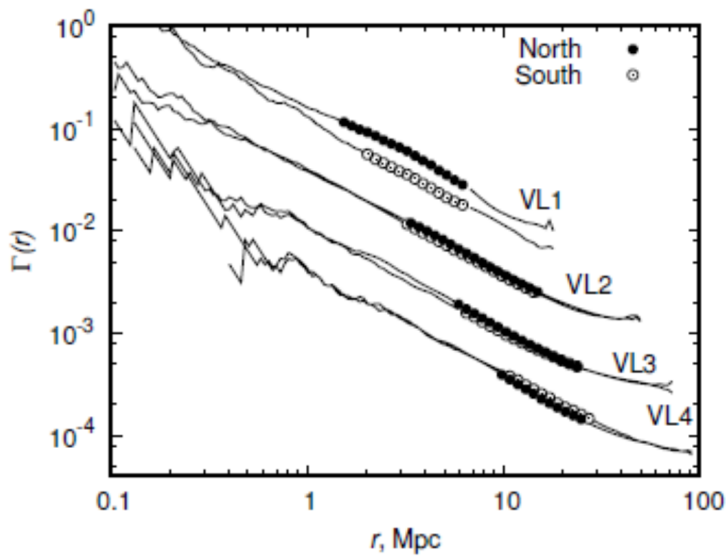

**Fig. 6.** *Conditional density for VL samples of 2MRS galaxies. The large dots mark the conditional density values in the scale length interval $r > R_{near}$ and $v(r) > 50\%$.*

The behavior of conditional density on short scale lengths is distorted because of the finite average distance between the nearest neighbors, which increases from about 1.5 Mpc for VL0 to 10 Mpc for VL3. Numerical simulations involving fractal structures of different dimensions showed that the correct slope of conditional density at small scale lengths can be seen starting from test sphere radii $r \sim 0.1 R_{near}$. This power-law behavior of conditional density can be observed from scale lengths of about 0.1 Mpc.

On large scale lengths the behavior of conditional density is distorted because of the abrupt decrease of the number of independent test spheres due to the requirement that the sphere should be fully submerged in the sample volume. This restriction begins to have a significant effect starting from scale lengths $r \sim 10$ Mpc for VL0 sample and amounts to $r \sim 70$ Mpc for VL3 sample. Such a behavior of conditional density is consistent with the extension of the true slope $\gamma = 1$ up to scale lengths of about 100 Mpc.

For each sample we identified the scale lengths that exceed the distance to the nearest neighbor in the sample considered such that at the same time $v(r) > 50\%$. The fact that the corresponding plots are parallel is a result of great importance, which follows form a comparison of $\Gamma$ values for samples with different threshold luminosity. This result suggests that the power-law slope (i.e., the character of clusterization) is independent of the galaxies IR luminosity interval considered.

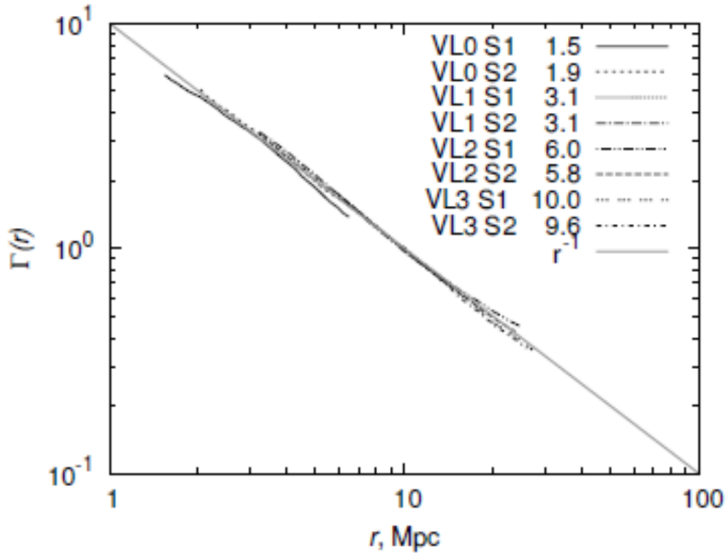

**Fig. 7.** *Conditional density for VL samples of 2MRS galaxies in the selected interval of scale lengths and normalized to 10 Mpc. The numbers in the legend give the average distance to the nearest neighbor.*

Figure 7 shows the conditional density (normalized so that $\Gamma(r = 10) = 1$) for VL samples limited to the scale length considered. Explicit fit of the conditional density for each sample yields an average exponent of $\gamma\_ave = 1.03 \pm 0.05$. Note that at largest scales of each VL sample the formal slope of conditional density has smaller values due to distortion by the sample border effects. E.g. for scale lengths $25 < r < 60$ Mpc in VL2 and and $40 < r < 80$ Mpc VL3 we get $\gamma\_ave = 0.46 \pm 0.08$. However, as we pointed out above, the effect of limited volume becomes important at these scale lengths. Furthermore, the conditional density for the corresponding samples behaves identically in the northern and southern hemispheres, which is indicative of the slope distortion due to the limited depth of the sample. We can therefore conclude hat given the distortions on short and long scale lengths, the observed power-law slope of conditional density is quite consistent with the universal slope $\gamma \approx 1$ over the entire scale's interval $0.1 < r < 100$ Mpc.

## 5. Simulated distributions of galaxies

### 5.1. Artificial Stochastic Fractal Distributions

Our simplest examples of a simulated distribution of galaxies that model the statistical properties of the real distribution were Cantor's stochastic structures with different fractal dimensions. Analysis of simulated distributions is important for studying the distorting factors that affect the measured slope of the power-law conditional density. We considered the three-dimensional conditional density for galaxies of the 2MRS catalog only in volume-limited samples and therefore had no to model the luminosity function. For an analysis of Cantor's stochastic distributions with the allowance for the luminosity function see [31].

As an example we consider Cantor's stochastic distribution with dimension $D = D_{crit} = 2$. Note that we consider a fractal dimension close to that observed in the 2MRS survey, however, the simulated distribution employed is not an exact model of the distribution of galaxies. The point is that fractal dimension alone cannot uniquely characterize the distribution of objects in space and an adequate model should also take into account the lacunarity of the fractal structure.

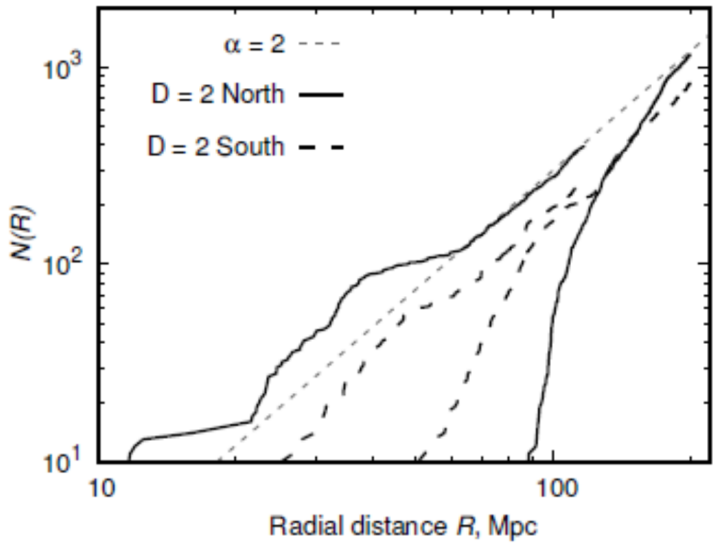

**Fig. 8.** *Radial distribution for a realization of Cantor's distribution with dimension D = 2 in two independent hemispheres is similar to that considered for VL2 sample of 2MRS galaxies.*

Figures 8 and 9 show the $N(R)$- and $SL(R, r)$-statistics for a realization of Cantor's stochastic fractal distribution, which reproduces S1 VL2 sample of the 2MRS catalog in terms of the geometry and the number of points. Figure 10 shows the results of the computation of conditional density for three independent generations of Cantor's distribution with a fractal demension of $D = 2$ in the full sphere of radius $R = 200$ Mpc and in two half-spheres. To compare the slope values $\Gamma(r)$, we multiplied the conditional density for different generations by coefficients equal to multiples of 10.

The measured slopes of conditional density for different generations are equal to $\gamma = 0.995 \pm 0.002$, $\gamma = 0.990 \pm 0.002$, and $\gamma = 1.009 \pm 0.003$. Deviations from the power-law relation at the shortest and longest scale lengths are caused by the distorting factors due to the finite size of the sample.

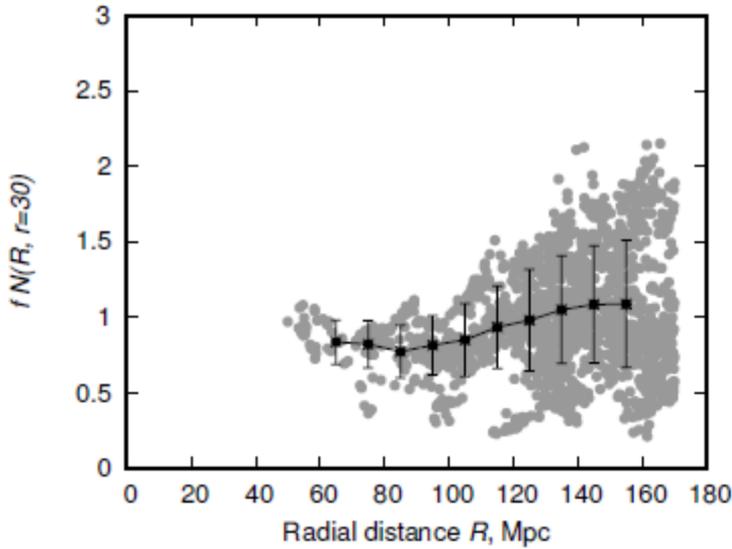

**Fig. 9.** *Normalized SL statistics (r = 30 Mpc) for a single realization of Cantor's distribution with dimension D = 2. The average increases with increasing distance.*

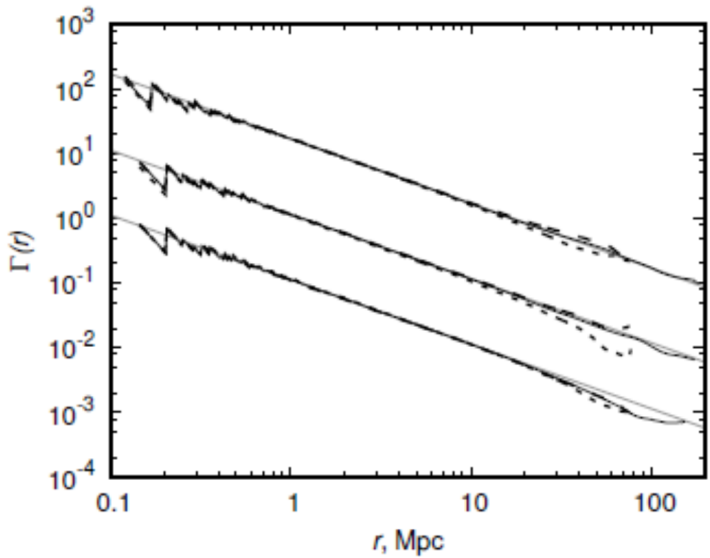

**Fig. 10.** *Conditional density of three generations of Cantor's distribution with fractal dimension D = 2 in the full sphere (the solid lines) and in the half-spheres (the dashed lines) of radius R = 200 Mpc. The slopes Γ(r) for different generations are multiplied by multiples of 10. The gray lines have the slope of γ = 1.*

### 5.2. Distribution of galaxies in Millennium catalog

We also analyzed the distribution of galaxies of the semianalytic distribution based on Millennium numerical simulation [32]. Galaxies of this catalog fill a cube of side 500 Mpc/h (where *h* is the dimensionless factor, which is equal to unity for the Hubble constant value of *H* = 100 km/ s Mpc). For further use we drew a subsample of this catalog by selecting only galaxies located in the sphere of radius 200 Mpc/*h* centered on the galaxy closest to the center of the cube. We then placed the observer inside this galaxy and used velocity data to create a new catalog containing distances of galaxies from the observer measured in the redshift space and computed with parameter *h* = 0.7, which corresponds to the Hubble constant value of *H* = 70 km/s Mpc, which we used when analyzing the 2MRS catalog.

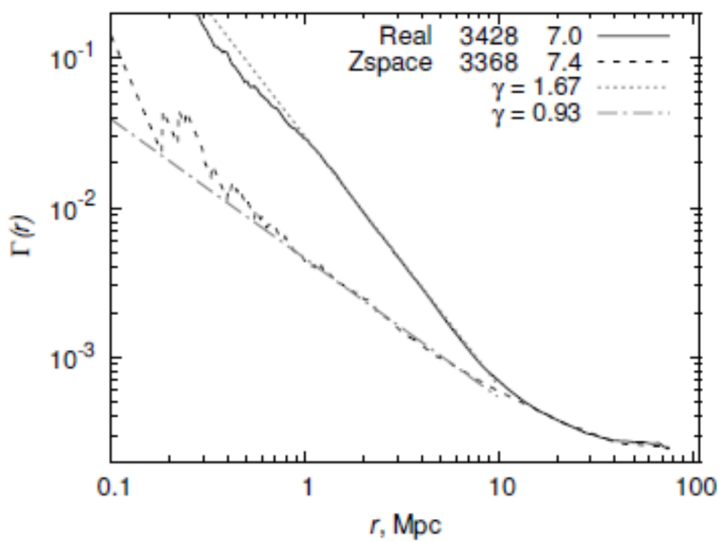

**Fig. 11.** *Conditional density of Millennium catalog galaxies in a sample similar to S1VL2 as a function of scale length in real and z space. The slopes are estimated in the 1 < r < 10 Mpc interval. Both curves are identical for scales   r > 10 Mpc.*

The important difference between the distribution of galaxies with and without peculiar velocity corrections applied is evident from Fig. 11. The exponent of the conditional density is equal to $\gamma = 1.669 \pm 0.005$ and $\gamma = 0.930 \pm 0.006$ in real and redshift space, respectively. Thus in the LCDM model the "true" distribution of galaxies is completely hidden by the dispersion of peculiar velocities and becomes uniform starting from scale lengths 10–30 Mpc. The conditional densities of the 2MRS and Millenium catalogs also differ by their slopes and this difference becomes apparent over a large interval of scale lengths.

Figure 12 shows the SL statistics of Millennium catalog for a sample with a geometry and density similar to those of S1VL2 sample of 2MRS catalog.

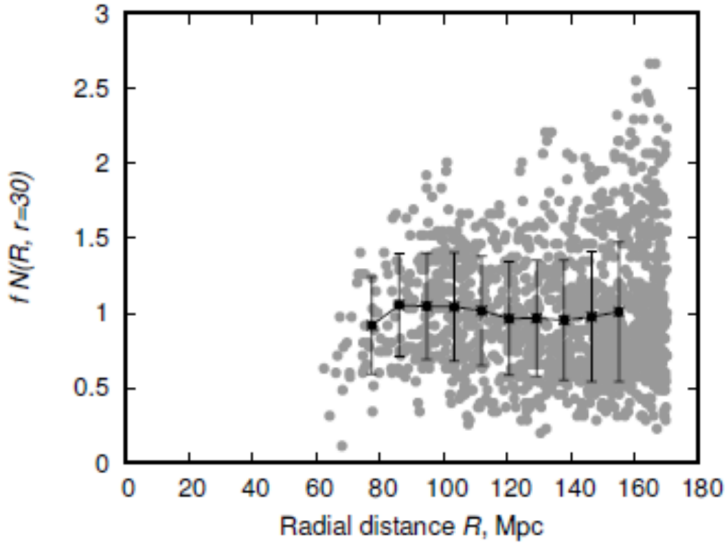

**Fig. 12.** *Normalized SL statistics for spheres of radius r = 30 Mpc in a sample of Millennium catalog similar to sample VL2 in the northern 2MRS hemisphere in terms of geometry and density. The average is constant with increasing distance.*

## 6. Results and discussion

According to [23], the aim of the 2MRS galaxy redshift survey is to "tomap the full three-dimensional distribution of galaxies in the nearby Universe". The 2MRS redshift survey of 2MASS galaxies has unique properties compared to optical SDSS surveys.

First, the samples of the 2MRS survey studied here uniformly cover 90% of the entire sky. Spherical geometry of the survey fundamentally expands the geometric boundaries of the samples and ensures the fulfillment of direction-averaging conditions. Second, interstellar extinction effects in the near infrared ($K$-band) 2MASS survey are one order of magnitude weaker than in the optical part of the spectrum. Furthermore, in the infrared the $k$ corrections and evolutionary corrections inside the sample volume are insignificant. The 2MRS survey is 97% complete down to the limiting magnitude of $Ks = 11.75$. Third, the spectral energy distributions of most of the galaxies have a peak in the near infrared and $K$-band luminosities are good approximations for the baryonic mass because at these wavelengths the "stellar mass to luminosity" ratio is almost constant for galaxies of different types. This circumstance allows one to analyze the space distribution of the entire population of galaxies as a whole [23].

We used these unique properties of the 2MRS galaxy redshift survey and the method of complete correlation functions to assess the nature of the observed global distribution of galaxies in the Local universe. The $N(R)$ statistics shows a large scatter in the behavior, because it is calculated for only one data point, where the observer is located. The SL($R, r$) statistics shows similar increasing behavior in the case of 2MRS and Cantor's stochastic fractals having the critical fractal dimension $D_{cdit} = 2$. However for the *Millennium* sample the SL statistics is constant with increasing distance.

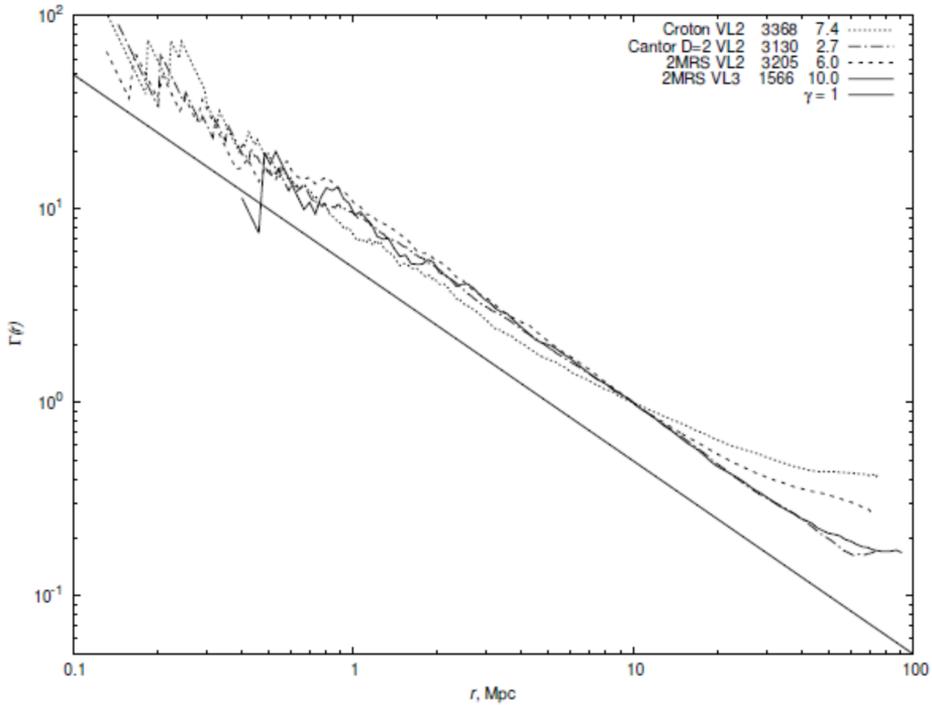

**Fig. 13.** *Comparison of the conditional density for VL samples of the 2MRS catalog and samples of identical geometry from simulated catalogs.*

Note that the complete correlation function $\Gamma(r)$ (conditional density) has an important advantage over reduced correlation function $\xi(r)$ (Peebles' two-point correlation function) in that the computation of conditional density requires no assumption about the uniformity of space distribution of galaxies analyzed. A comparison (Fig. 13) of complete correlation functions for samples of 2MRS galaxies with the correlation functions of simulated distributions shows that:

- The observed global space distribution of 2MRS galaxies, determined with taking into account the factors that distort conditional density at the largest scale lengths, can be described by the power-law complete correlation function of the form $\Gamma(r) \sim r^{-\gamma}$ with the slope $\gamma \approx 1$ over a wide interval of scale lengths spanning from 0.1 to 100 Mpc and is consistent with simple stochastic fractal model with the critical fractal dimension of $D \approx D\_{crit} = 2$.
- The power-law slope $\gamma \approx 1$ of conditional density $\Gamma$ does not depend on the galaxy luminosity interval as it follows from the parallel behavior of conditional density for different VL samples.

- According to the prediction of the LCDM model (catalog [32]), conditional density has a power-law form $\Gamma(r) \sim r^{-\gamma}$ in the scale interval $0.1 < r < 20$ Mpc and becomes uniform starting from 20 Mpc. Note that slope $\gamma$ has two significantly different values: $\gamma = 1.67$ for real space and $\gamma = 0.832$ for redshift space. This is due to the dispersion of peculiar velocities of galaxies. The scale length is the order of 10 Mpc so far separates nonlinear clustering mode from linear perturbation growth in a uniform medium.

- The observed extension (up to 100 Mpc) of universal fractal dimension $\gamma \approx 1$ from the domain with nonlinear clusterization ($r < 10$ Mpc) into the domain of linear growth of structures ($r > 10$ Mpc) is a sign of "cosmic conspiracy", where the effect of the supposed distribution of peculiar velocities does not change the slope of the complete correlation function (which is equal to $\gamma \approx 1$) independently on whether these scale lengths belong to nonlinear stage of clusterization or to linear mode of perturbation growth.

Note that the presence of observed structures in the Local Universe with the sizes of several hundred Mpc is naturally consistent with the global stochastic fractal distribution of galaxies at the scale interval from 0.1 Mpc up to 100 Mpc.

We point out in conclusion that there are general physical arguments in favor of the fundamental role of the critical fractal dimension $D_{crit} = 2$ in different physical systems with self-gravity [2, 17, 21, 22]. Another important problem of theoretical analysis of the space distribution of galaxies is the applicability of classical regular description of hydrodynamics relations connecting peculiar velocities with mass density fields by smooth analytical functions, whereas new mathematical methods of analysis are required in the case of discrete stochastic hierarchical distributions [16]. The capability of the theory to answer many questions regarding the large-scale distribution of galaxies will depend on the progress in the field of galaxy distance measurements using redshift-independent methods. This will make it possible to measure the real peculiar velocities of galaxies and determine their effect on the measured correlation properties of the space distribution of galaxies in the Local Universe.

## Acknowledgements

This work was supported by the St.-Petersburg State University (grant No. 6.38.18.2014).